\documentclass[twocolumn,showpacs,prl,amsmath,amssymb,floatfix,superscriptaddress]{revtex4-1}
\usepackage{bm,color,amsmath,amssymb,mathrsfs,latexsym,graphicx,psfrag,float}

\usepackage{graphicx,color}
\usepackage[bookmarks]{hyperref}
\usepackage{amssymb}

\definecolor{myblue}{rgb}{0,0,0.75}

\newcommand{\be}{\begin{equation}}
\newcommand{\ee}{\end{equation}}
\newcommand{\eq}[1]{Eq.~(\ref{#1})}

\begin{document}
\title{Quenching a Quantum Critical State by the Order Parameter: Dynamical Quantum Phase Transitions and Quantum Speed Limits}
\author{Markus Heyl}
\affiliation{Physik Department, Technische Universit\"at M\"unchen, 85747
Garching, Germany}
\affiliation{Max Planck Institute for the Physics of Complex Systems, Dresden 01187, Germany}

\begin{abstract}

Quantum critical states exhibit strong quantum fluctuations and are therefore highly susceptible to perturbations. In this work we study the dynamical stability against a sudden coupling to these strong fluctuations by quenching the order parameter of the underlying transition. Such a quench can generate superextensive energy fluctuations. This leads to a dynamical quantum phase transition (DQPT) with nonanalytic real-time behavior in the resulting decay of the initial state. By establishing a general connection between DQPTs and quantum speed limits, this allows us to obtain a yet unrecognized quantum speed limit with unconventional system size dependence. These findings are illustrated for the one-dimensional and the infinitely-connected transverse-field Ising model. The main concepts, however, are general and can be applied also to other critical states. An outlook is given onto the implications of the superextensive energy fluctuations on potential restricted thermalization despite of nonintegrability.

\end{abstract}

\maketitle
\date{\today}

\emph{Introduction.--} Systems in the vicinity of quantum phase transitions experience strong quantum 
fluctuations and correlations which gives them also interesting dynamical properties~\cite{Hohenberg1977}. This includes 
critical slowing down~\cite{Hohenberg1977} or the creation of defects in the context of the Kibble-Zurek mechanism~\cite{Kibble1976,Zurek1985,Zurek1996} when 
slowly sweeping through a quantum critical point~\cite{Dziamarga2010,Gritsev2010}. While this has led to a comprehensive understanding of the long-time dynamics in the vicinity of critical points, here we concentrate onto the equally challenging regime of transient nonequilibrium response. 

In this work the transient dynamics of a quantum critical state is studied after a sudden coupling to its strong quantum fluctuations by quenching the order parameter. 
As the main result, it is found that such a quench induces a dynamical quantum phase transition (DQPT)~\cite{Heyl2013a} yielding nonanalytic behavior during quantum real-time evolution. 
In particular, the strong quantum fluctuations in the initial critical state lead to a critical time for the dyamical transition that turns out to exhibit an unconventional system-size dependence. Specifically, the critical time vanishes in the thermodynamic limit implying a breakdown of time-dependent perturbation theory immediately after the quench. 
Furthermore,  this breakdown implies that the initial critical state becomes orthogonal to itself after a short time evolution signaling optimal distinguishability of the two states.
This observation allows us to establish a general connection between DQPTs and quantum speed limits which has profound implications for the studied dynamics.
While the main ideas are illustrated using two paradigmatic model systems of quantum criticality, the one-dimensional and the infinitely-connected transverse-field Ising model, the concepts are general and also apply to other systems~\cite{supp}.

\emph{Setup.--} Consider a system initially prepared in a pure state $|\psi_0\rangle$ which in the following is taken to be the ground state of a Hamiltonian $H_0$ at its critical point.  Upon suddenly switching a parameter $h$ in the Hamiltonian $H_0 \mapsto H = H_0 + h \mathcal{O}$ (here $\mathcal{O}$ will be chosen as the order parameter of the transition), the decay of the initial state can be characterized through the Loschmidt amplitude: 
\be 
\mathcal{G}(t) = \left\langle \psi_0 \right| e^{-iHt} \left| \psi_0 \right\rangle.
\label{eq:LoschmidtAmlitude}
\ee 
Objects of the structure of $\mathcal{G}(t)$ appear  as quantifiers for the stability of quantum states during unitary evolution in many contexts such as the Schwinger mechanism in high-energy physics~\cite{Schwinger1951,Martinez2016}, quantum chaos~\cite{Peres1984kq,Gorin2006dv}, or quantum speed limits~\cite{Mandelstam,Levitin,Campo,Deffner}.

Moreover, Loschmidt amplitudes play a central role in the theory of dynamical quantum phase transitions (DQPTs)~\cite{Heyl2013a} which has developed into an emerging prototype of phase transitions far from equilibrium experiencing significant interest~\cite{Pollmann2010dv,Takahashi2012,Karrasch2013,Andraschko2014,Kriel2014,Hickey2014,Vajna2014,Canovi2014fo,Heyl2014,Vajna2015,Schmitt2015,Heyl2015dq,Budich2016,Zvyagin2015,Divakaran2015,Sharma2015,Abeling2016,James2015,Huang2016,Sharma2016,Zunkovic2016,Flaeschner2016,Jurcevic2016}. Very recently, DQPTs have been observed experimentally for the first time~\cite{Flaeschner2016,Jurcevic2016}. As opposed to conventional equilibrium phase transitions that are driven by control parameters such as temperature or pressure, DQPTs occur during nonequilibrium quantum real-time evolution with Loschmidt amplitudes becoming nonanalytic at critical times. DQPTs have been identified in various models~\cite{Pollmann2010dv,Takahashi2012,Karrasch2013,Andraschko2014,Kriel2014,Hickey2014,Vajna2014,Canovi2014fo,Heyl2014,Vajna2015,Schmitt2015,Heyl2015dq,Budich2016,Zvyagin2015,Divakaran2015,Sharma2015,Abeling2016,James2015,Huang2016,Sharma2016,Zunkovic2016,Flaeschner2016,Jurcevic2016} and recently substantial progress has been achieved for topological systems~\cite{Vajna2015,Budich2016,Huang2016,Flaeschner2016}, by identifying dynamical order parameters~\cite{Budich2016,Flaeschner2016}, scaling, universality~\cite{Heyl2015dq}, or robustness~\cite{Karrasch2013,Kriel2014,Sharma2015}. It is one purpose of this work to point out an interesting connection to another important concept in quantum physics -- quantum speed limits.

\emph{Model -- }In the following, the main ideas will be illustrated using the one-dimensional transverse-field Ising chain:
\be
	H_0(g) = -J \sum_{l=1}^{N-1} S_l^z S_{l+1}^z - g \sum_{l=1}^N S_l^x.
\ee
Here, $S_l^\alpha$ are spin-$1/2$ operators with $\alpha=x,y,z$, $l=1,\dots,N$, and $N$ the total number of lattice sites. Open boundary conditions are used in the following. The quantum critical point for this model is located at $g/J = 1/2$ separating an ferromagnetic phase ($g/J<1$) from a paramagnetic phase ($g/J>1$). The order parameter of the transition is the magnetization $\mathcal{M}  = \sum_l S_l^z$, which we therefore take as our perturbation $\mathcal{O}=\mathcal{M}$ for the quench. In the remainder we choose units where $\hbar=1$ and the zero of energy such that $|\psi_0\rangle$ has vanishing expectation value with respect to $H_0=H_0(g/J=1/2)$. At the end of the article, we will also discuss another paradigmatic model of phase transitions, the infinitely-connected transverse-field Ising model.

\emph{Cumulant generating function of energy.--}
The Loschmidt amplitude $\mathcal{G}(t)$ is the Fourier transform of the energy (work) distribution function~\cite{Talkner2007aw,Silva2008gj} and thus 
\be 
K(t) = -\log\left( \mathcal{G}(t) \right) = -\sum_{l=1}^\infty \frac{1}{l!} \kappa_l (-it)^l, 
\label{eq:cumulantGeneratingFunction}
\ee 
is the respective cumulant generating function with $\kappa_l$ denoting the cumulants. For noncritical states, the $\kappa_l$ are extensive and we have that $\mathcal{G}(t)$ satisfies a large-deviation scaling~\cite{Silva2008gj,Heyl2013a,Gambassi2012a} with $\mathcal{G}(t) = \exp[-Nk(t)]$. Thus, $K(t) = -N k(t) $ with $k(t)$ intensive, $N=L^d$ the system size and $L$ denotes the linear extent of the system and $d$ the dimension. If the problem at hand is perturbative at short times we have that $k(t) = i\epsilon t + \Delta \epsilon^2 t^2/2 + \mathcal{O}(t^3)$. Here, $\epsilon = N^{-1} \kappa_1 = N^{-1} \langle H \rangle = N^{-1} \langle \psi_0 | H | \psi_0 \rangle$ the mean energy density and $\Delta \epsilon^2 = N^{-1} \kappa_2 = N^{-1} [ \langle H^2 \rangle - \langle H \rangle^2]$ the energy fluctuation density in the initial state. 

The rate function $K(t)$ can become nonanalytic as a function of time which is the defining feature of the anticipated DQPTs~\cite{Heyl2013a}. This is possible because, formally, Loschmidt amplitudes resemble conventional equilibrium partition functions at complexified parameters. 
Specifically, objects of the structure $Z_B = \langle \psi_1 | e^{-RH} | \psi_2 \rangle$ appear as boundary partition functions in equilibrium where the states $|\psi_{1/2}\rangle$ encode the boundary conditions on two ends of a system and $R$ denotes their distance~\cite{LeClair1995ho}.  Replacing $R \to it$ and $|\psi_{1/2}\rangle \to | \psi_0\rangle$ Loschmidt amplitudes can be thought of as a Wick-rotated partition function. Analogously, the initial state $|\psi_0\rangle$ in the dynamical problem can be identified as a boundary condition in time. While the notion of dynamical phase transitions also appears in other contexts~\cite{Eckstein2009,Schiro2010gj,Sciolla2010jb,Hedges2009kn,Garrahan2010xw,Diehl2010,Mitra2012}, in the following the definition in terms of Loschmidt amplitudes will be adopted here.

\emph{Divergent energy fluctuations and entanglement.--}
When considering initial quantum critical states, the extensivity of the cumulant generating function $K(t)$ can be lost.  While $\epsilon=0$, since the order parameter has vanishing expectation value at the critical point, standard scaling theory implies that energy fluctuations $\Delta E^2 = L^d \Delta \epsilon^2$ can become superextensive:~\cite{supp}
\be
\Delta E^2 \propto L^{2d-2\Delta_\mathcal{O}},
\label{eq:divergentFluctuations}
\ee
when $d>2\Delta_\mathcal{O}$. Here, $\Delta_\mathcal{O}$ denotes the scaling dimension of the operator $\mathcal{O}$. If $d<2\Delta_\mathcal{O}$ the critical fluctuations of the order parameter do not overcome the nonuniversal contributions from short-range correlations which contribute conventional extensive energy fluctuations. Thus, $\Delta E^2$ does not become superextensive in that case. For $d=2\Delta_\mathcal{O}$ also logarithmic corrections are possible.

This potential divergence of energy fluctuations roots in the strong quantum correlations at a critical point because $\Delta \epsilon^2 \propto  N^{-1} \sum_{lm=1}^N \langle \mathcal{O}_l \mathcal{O}_m \rangle$ measures the order parameter structure factor. Notice that there is an interesting connection to divergent entanglement at quantum phase transitions for operators $\mathcal{O}$ of the structure $\mathcal{O} = \sum_{l=1}^N f_l \sigma_l^\alpha $, $\alpha=x,y,z$, with $\sigma_l^\alpha$ Pauli matrices and $f_l=1$ or $f_l = (-1)^l$. Then, $f_Q = (4N)^{-1} \sum_{lm=1}^N \langle \mathcal{O}_l \mathcal{O}_m \rangle$ is a quantum Fisher information and therefore a witness for multipartite entanglement~\cite{Hyllus2012,Toth2014,Strobel2014,Hauke2016}. In other words, divergent energy fluctuations can be associated with divergent entanglement in the initial state.  While an entanglement witness in general only bounds entanglement and cannot be considered an entanglement measure or monotone, the quantum Fisher information has turned out to be a valuable quantifier for entanglement at quantum phase transitions~\cite{Hauke2016}.

\begin{figure}
\centering
\includegraphics[width=\columnwidth]{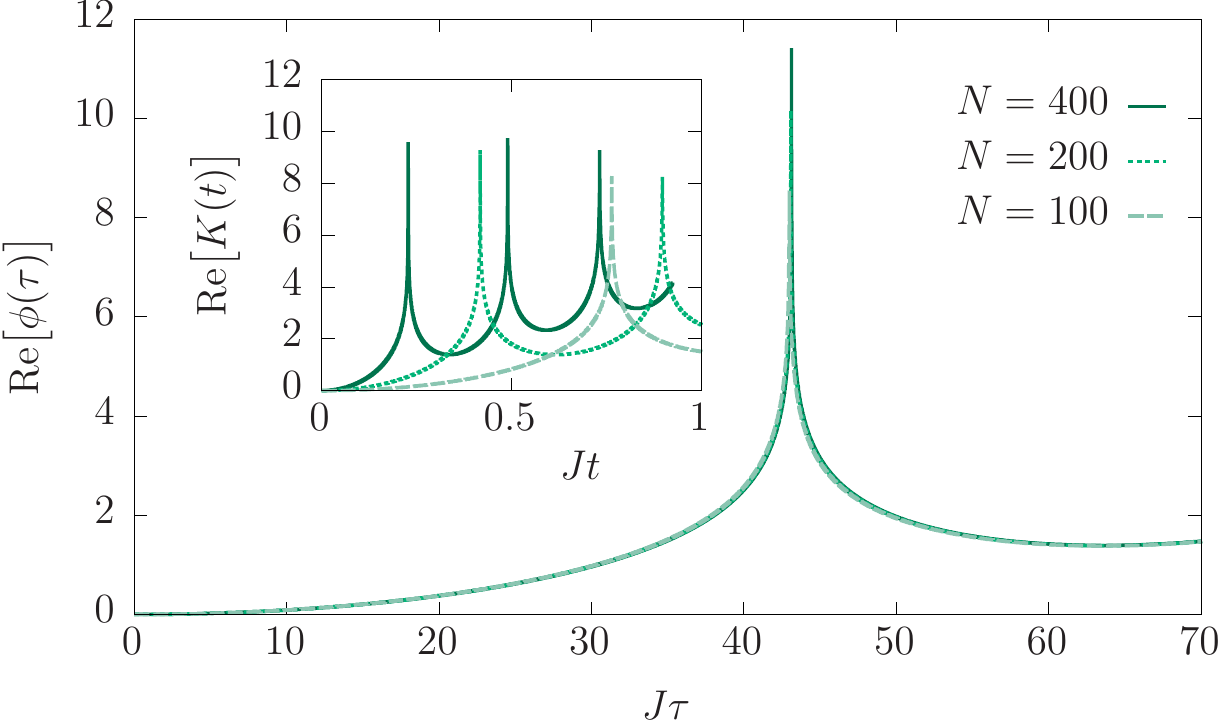}
\caption{Dynamics of the rescaled cumulant generating function $\phi(\tau)$ in the one-dimensional transverse-field Ising chain for different system sizes $N$ at a longitudinal field strength $h/J=0.1$. For comparison, the inset shows the cumulant generating function $K(t)$ before rescaling.}
\label{fig:phi_Ising1D}
\end{figure}

In the presence of the strong energy density fluctuations, see \eq{eq:divergentFluctuations}, the cumulant generating function $K(t)$ cannot be extensive at short times 
as we have for noncritical states. In contrast, it is one main result of this work that for the considered models $K(t)$ satisfies the following general functional form:
\be
	K(t) = L^{a} \phi\left( t L^b  \right),
	\label{eq:defPhi}
\ee
In Fig.~\ref{fig:phi_Ising1D} one can see $\phi(\tau)$ for the quench in the Ising chain. One obtains an excellent collapse of the data for different system sizes with the exponents $a=0$, $b=7/8$ and we have defined $\tau = t L^b$. From the numerical data the exponent $b$ can be determined from the system-size dependence of the first peak whereas the exponent $a$ by performing a scaling collapse. For the presented data this gives $b=0.875(2)$ and $a=0.001(5)$ consistent with $b=7/8$ and $a=0$. By varying the symmetry-breaking field $h$ the main features do not change except that the time scales become larger for decreasing $h$. This data has been obtained using the iTensor library~\cite{iTensor} with a Trotter time step of $\Delta t =10^{-4} L^{-7/8}$ and bond dimension $\chi=200$. We have checked that the data has converged both concerning $\chi$ as well as $\Delta t$. 

The exponents $a$ and $b$ can be constrained by matching the general form of $K(t)$ to the expansion at small times. Because $K(0) = \partial_t K(t=0) = 0$ we have that $\Delta E^2 t^2 \propto L^{a+2b} t^2$ together with Eq.~(\ref{eq:divergentFluctuations}) and thus
\be
	a+2b = 2d-2\Delta_\mathcal{O}.
	\label{eq:exponent_constraint}
\ee
While this constraint does not uniquely determine the individual exponents, knowing either $a$ or $b$, however, is sufficient. Importantly, the exponents $a=0$ and $b=7/8$ are exactly compatible because for the transverse-field Ising chain we have that $d=1$ and $\mathcal{O}=1/8$.

\begin{figure}
\centering
\includegraphics[width=\columnwidth]{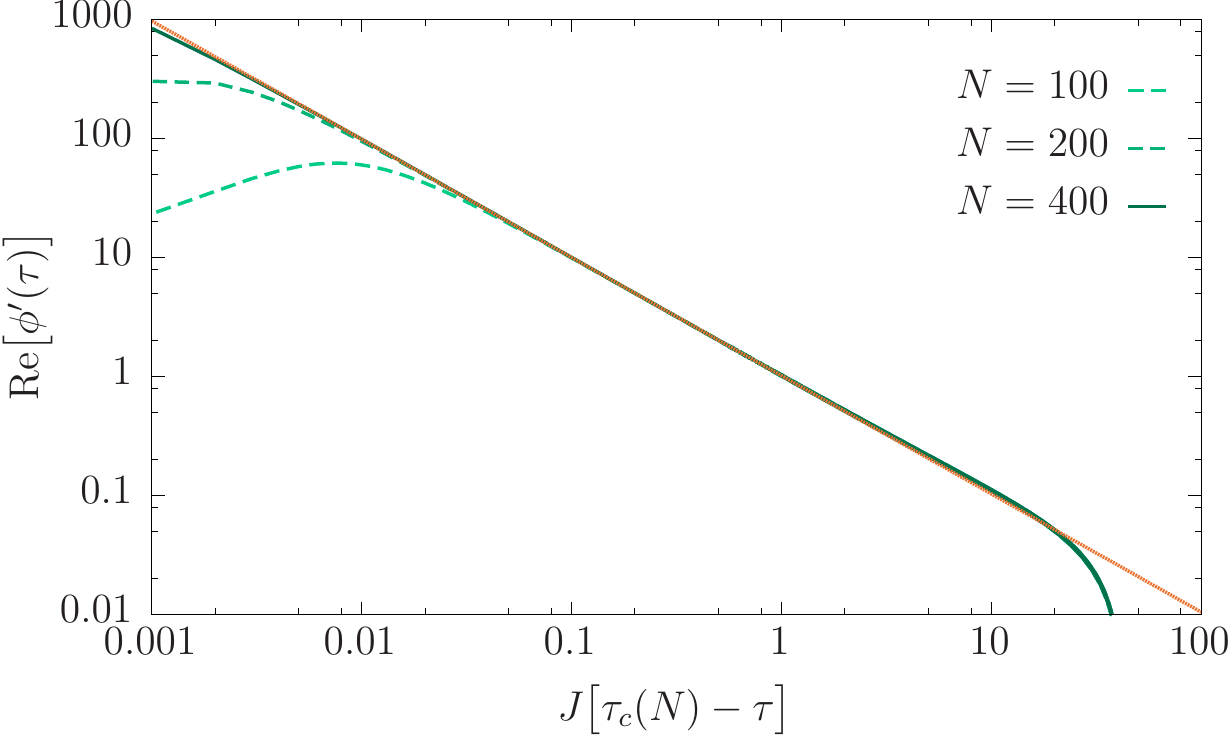}
\caption{Dynamical quantum phase transition in the one-dimensional transverse-field Ising chain for the same parameters as in Fig.~\ref{fig:phi_Ising1D}. The numerical data for the derivative $\phi'(\tau) = d \phi(\tau)/d\tau$ indicates a power-law divergence. An algebraic fit $|\tau_c(N)-\tau|^{-\alpha}$ to the curve for $N=300$, with $\tau_c(N)$ the finite-size pseudo-critical point, gives an exponent $\alpha = 0.98(2)$.}
\label{fig:Ising1D_nonanalytic}
\end{figure}

\emph{Dynamical quantum phase transitions.--}
While for short times the initial increase of $\phi(\tau)$ is still quadratic one observes a prominent peak at larger times. In Fig.~\ref{fig:Ising1D_nonanalytic} numerical evidence is provided that this peak develops into a nonanalytic structure in the thermodynamic limit which is the defining feature of a DQPT~\cite{Heyl2013a}. Specifically, we plot $\phi'(\tau) = \partial_\tau \phi(\tau)$ showing evidence for a power-law behavior in the vicinity of the  sharp structure. In particular, we find from a power-law fit to the data that $\phi'(\tau) \sim (\tau_c-\tau)^{-\alpha}$ with $\alpha = 0.98  (2)$. The system-size dependent pseudo-critical time $\tau_c(N)$ has been determined by the local maximum of $\phi(\tau)$ at a given $N$.

The emergence of a DQPT at a time $\tau_c$ implies the breakdown of time-dependent perturbation theory in analogy to the breakdown of high-temperature series expansions at equilibrium thermal phase transitions. As will be shown later, this has important consequences for quantum speed limits~\cite{Mandelstam,Levitin,Campo,Deffner}. While still for any system of finite size we can use \eq{eq:cumulantGeneratingFunction} to expand $K(t)  = -\Delta E^2 t^2/2 + \mathcal{O}(t^4)$, the radius of convergence $t^\ast(L)$ of this series is necessarily limited by the critical $\tau_c$:
\be
    t^\ast(L) = \tau_c L^{-b},
    \label{eq:radiusOfConvergence}
\ee
with $b$ given by Eq.~(\ref{eq:defPhi}). Importantly, $t^\ast(L)$ vanishes in the thermodynamic limit which is different from previously studied cases where the critical times of DQPTs have always been found to be independent of system size.

\emph{Quantum speed limits.--}
Quantum speed limits give general bounds on the time scale for how fast quantum states evolve in real-time dynamics~\cite{Mandelstam,Levitin,Campo,Deffner}. This need not be the speed of change for local observables, but rather quantifies at which point in time a time-evolved state becomes distinguishable from the initial one. Besides of setting fundamental limits for the dynamics in closed~\cite{Mandelstam,Levitin} and open~\cite{Campo,Deffner} quantum systems, quantum speed limits also have applications in optimal control theory~\cite{Caneva} and are believed to be important for many quantum technologies such as quantum metrology as has been argued, for example, in Refs.~\cite{Campo,Deffner}.

Optimal distinguishability of two quantum states is achieved when they are orthogonal. In terms of the dynamical problem this implies a vanishing overlap or Loschmidt amplitude, see \eq{eq:LoschmidtAmlitude}.  The Mandelstam-Tamm bound~\cite{Mandelstam} limits the time scale $T$ necessary for a state becoming orthogonal to itself under coherent real-time evolution with a time-independent Hamiltonian by:
\be
      T \geq \frac{\pi}{2\Delta E}.
\ee
Although it is straightforward to imagine that states can become orthogonal during time evolution for small systems, e.g., a single spin performing Larmor precession in a magnetic field, for a many-body system it appears unlikely in general that Loschmidt amplitudes can vanish. To see this, consider the spectral representation of the Loschmidt amplitude:
\be
	\mathcal{G}(t) = \sum_{\nu} \left| \langle \phi_\nu | \psi_0 \rangle \right|^2 e^{-iE_\nu t},
\ee
with $|\phi_\nu\rangle$ denoting the eigenstates of the final Hamiltonian $H$ and $E_\nu$ the corresponding energies. As one can see from this formula, taking a system of finite size, exact zeros of $\mathcal{G}(t)$ require a fine-tuned ``phase condition''~\cite{Huang2016} on all $e^{-iE_\nu t}$ to exactly cancel all the involved terms which is, in general, not possible.

The situation, however, changes for large many-body systems. As anticipated before, the Loschmidt amplitude can be interpreted as a conventional partition function at complexified parameters. The complexification of parameters is important for the equilbrium theory of phase transitions which leads to the concept of Fisher~\cite{Fisher1967da} and Lee-Young zeros~\cite{YangLee}, a concept which can consequently also applied to $\mathcal{G}(t)$. Within this analogy, $\mathcal{G}(t)$ is determined by its zeros $z_n$ in the complex plane by extending $t\to z \in \mathbb{Z}$~\cite{Heyl2013a}: $\mathcal{G}(z) = e^{\mu(z)} \prod_{n} ( z-z_n \big)$ with $\mu(z)$ a smooth function. The singular contribution $K_s(t)$ to $K(t)$ is given by $K_s(t) = -\int dz \, \rho(z) \log( t-z )$ with $\rho(z) = \sum_n \delta(z-z_n)$ the density of zeros~\cite{Schmitt2015}. For a finite-size system these zeros are generically located on isolated points in the complex plane and require fine-tuning to lie exactly on the real-time axis because of the anticipated phase condition~\cite{Huang2016}. In the thermodynamic limit on the other hand the zeros accumulate to form lines or areas. Whenever such a line or area crosses the real-time axis, $K(t)$ becomes nonanalytic~\cite{Heyl2013a} as is the case at conventional equilibrium transitions~\cite{Fisher1967da,YangLee}.

The vanishing Loschmidt amplitude associated with these zeros implies that at a DQPT the initial and time-evolved state become optimally distinguishable. Thus, for the order parameter quench of the critical state we find that the time $T$ for distinguishability relevant for quantum speed limits is set by the DQPTs giving:
\begin{equation}
      T = t^*(L) = \tau_c L^{-b}.
\end{equation}
Notice that this close relationship between quantum speed limits and DQPTs is not just restricted to the present problem, but is rather general and not related to details of the studied model system.

While it has already been realized that entangled states can lead to an enhanced system-size dependent speed of evolution~\cite{Giovanetti2013,Batle2005,Borras2006,Zander2007}, it is important to emphasize that the origin for the time scale $t^\ast(L)$ reported in the present work is different in nature. This is because $t^\ast(L)$ does not estimate the short time evolution on the basis of the first few cumulants but rather the full radius of convergence which gives the profound connection to DQPTs.

\emph{Infinitely-connected Ising model.--}
After having discussed the main ideas, results for another paradigmatic model system for phase transitions will be presented, the infinitely-connected transverse-field Ising model:
\be
	H_0(h) = - \frac{J}{N} \sum_{l<m=1}^N S_l^z S_m^z - g \sum_{l=1}^N S_l^x,
\ee
which in contrast to the previous case also exhibits phase transitions at nonzero temperatures. This system has its quantum critical point at $g/J = 1$ separating a ferromagnetic phase ($g/J<1$) from a paramagnetic one ($g/J>1$). The order parameter of the transition is the magnetization $\mathcal{M}  = \sum_l S_l^z$, i.e., for the considered quench this implies $\mathcal{O}=\mathcal{M}$ at  $g/J=1$. This model is exactly solvable even by adding the symmetry-breaking order parameter because the Hamiltonian commutes with $\vec S^2 = \sum_{\alpha=x,y,z} S_\alpha^2 $ where $S_\alpha = \sum_l S_l^\alpha$. As a consequence, the Hamiltonian becomes block diagonal in the  eigenbasis of $\vec S^2$ where the  largest of these blocks has a dimension of only $N + 1$. Considering this largest block one can study substantially larger system sizes  of up to $N=3000$ spins.

\begin{figure}
\centering
\includegraphics[width=\columnwidth]{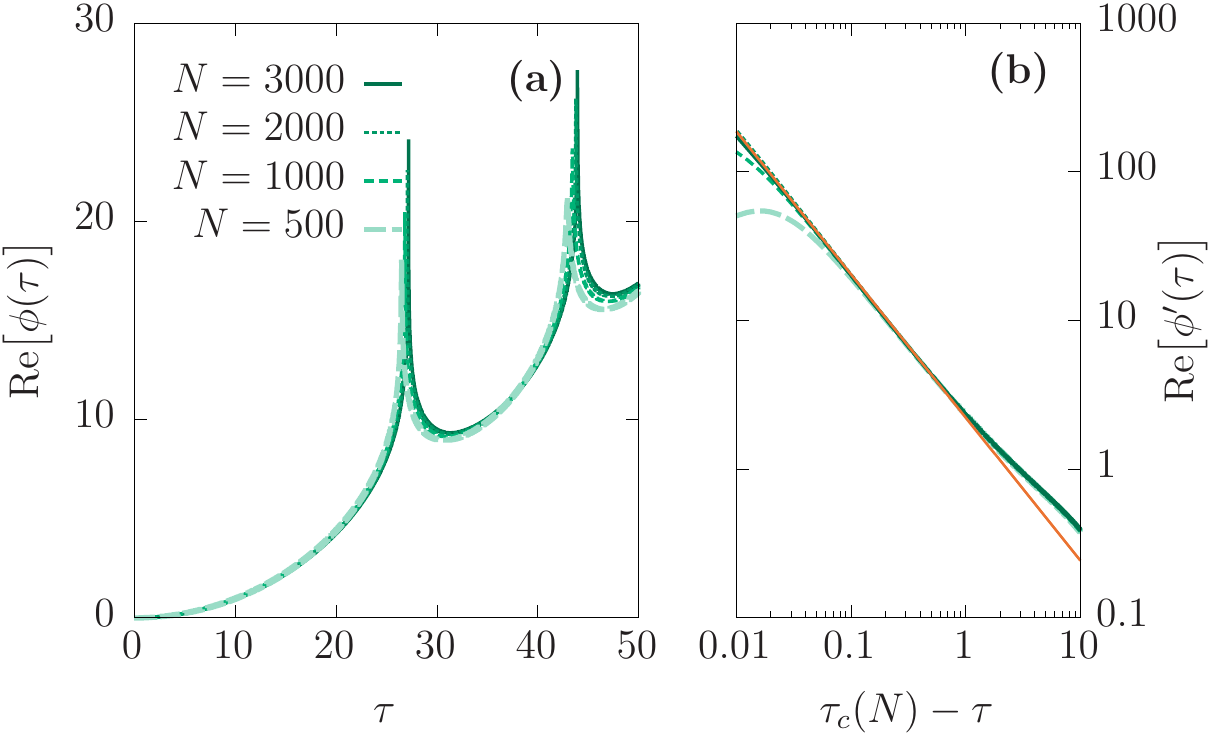}
\caption{Dynamical quantum phase transition for the infinitely-connected Ising model in a transverse field at a longitudinal field strength $h/J=0.2$. {\bf (a)} Rescaled cumulant generating function $\phi(\tau)$ for different system sizes $N$. {\bf (b)} Zoom onto the vicinity of the first peak located at $\tau_c(N)$ on a double logarithmic scale indicating again a algebraic divergence for increasing system size. From a power-law fit $|\tau_c(N)-\tau|^{-\alpha}$ to the data we find $\alpha = 1.00(5)$.
}
\label{fig:Ising_nonanalytic}
\end{figure}

In Fig.~\ref{fig:Ising_nonanalytic} the data collapse for the rescaled cumulant generating function $\phi(\tau)$ is shown for different system sizes. Again we find $a=0$ which using the constraint in Eq.~(\ref{eq:exponent_constraint}) implies $b=2/3$ because of the scaling dimension $\Delta_\mathcal{O}=1/3$ for the magnetization in the infinitely-connected Ising model~\cite{Botet1983}. In accord with the results obtained for the one-dimensional Ising chain, the derivative $\phi'(\tau)$ shows strong numerical evidence for a power-law divergence. From an algebraic fit $|\tau-\tau_c|^{-\alpha}$ to the data, we find that $\alpha=1.00(5)$.

\emph{Outlook.--}
In order to experimentally observe DQPTs and quantum speed limits for quenching a quantum critical state, it is first of all necessary to measure Loschmidt amplitudes. This is currently accessible in systems of trapped ions, where $\mathcal{
G}(t)$ has been recently measured~\cite{Martinez2016,Jurcevic2016}, or cold atoms in optical lattices where $\mathcal{G}(t)$ is also, in principle, experimentally feasible using a protocol~\cite{Daley2012} that has been recently implemented to measure entanglement properties in small systems~\cite{Islam2015me}. Moreover, the tomography technique proposed in Ref.~\cite{Hauke2014} and experimentally realized in Refs.~\cite{Flaeschner2016Science,Flaeschner2016} can be used to reconstruct Loschmidt amplitudes for noninteracting fermionic systems. In all these quantum-optical platforms, however, preparation of quantum critical ground states is challenging. One route towards generating such states is adiabatic state preparation which has already been used to tune noninteracting fermionic systems across a topological phase transition with sufficiently high fidelity~\cite{Aidelsburger2014mt,Jotzu2014}. The desired goal of creating a state close to the ground state at a critical point is thus within the scope of current experiments by stopping the sweep at the respective critical point.

Another interesting prospect of the present work is the question of thermalization in the long-time limit for the considered nonequilibrium quench protocol. The transverse-field Ising chain with longitudinal field is nonintegrable~\cite{Benenti2009} and thus expected to be thermalizing. The superextensive energy fluctuations are, however, not compatible with a thermal state. How these strong fluctuations influence the thermalization dynamics in the long-time limit is an interesting question for future work.

\begin{acknowledgments}
Discussions with Markus Schmitt are greatfully acknowledged. This research was supported by the Deutsche Akademie der Naturforscher Leopoldina under grant number LPDR 2015-01 and by the Deutsche Forschungsgemeinschaft via the Gottfried Wilhelm Leibniz Prize programme.
\end{acknowledgments}

\bibliography{literature}

%%%%%%%%%%%%%%%%%%%%%%%%%%%%%%%%%%%%%%%%%%%%%%%%%%%%%%%%%%%%%%%%%%%%%%%
%%%%%%%%%%%%%%%%%%%%%%%%%%%%%%%%%%%%%%%%%%%%%%%%%%%%%%%%%%%%%%%%%%%%%%%
%	Supplemental Material

\clearpage
\onecolumngrid
\setcounter{equation}{0}
\setcounter{figure}{0}

\begin{center}
{\bf \Large 
Supplemental Material to\vspace*{0.3cm}\\ 
\emph{Quenching a Quantum Critical State by the Order Parameter: Dynamical Quantum Phase Transitions and Quantum Speed Limits}
}
\end{center}

\vspace*{0.3cm}
{\center{
\hspace*{0.1\columnwidth}\begin{minipage}[c]{0.8\columnwidth}
In this supplement, further supporting material is provided on the scaling of energy fluctuations at a quantum critical point and numerical data for dynamical quantum phase transitions in the XXZ chain. 
\end{minipage}
}
}

\section{Scaling of the energy fluctuations}
\label{app:sec:scalingEnergyFluctuations}

In the following, the scaling of the energy fluctuations will be discussed using elementary scaling arguments that have been recently presented extensively in Ref.~\cite{Hauke2016}. In the remainder, we will follow closely this prescription. The object under consideration is the energy fluctuation density:
\be
	\Delta \epsilon^2 = \frac{1}{L^d} \sum_{lm=1}^N \langle \mathcal{O}_l \mathcal{O}_m \rangle,
\ee
which is nothing but the integrated order parameter correlations for $\mathcal{O} = \sum_l \mathcal{O}_l$. Here $L$ denotes the linear extent of the system and $d$ the dimension such that the total number of lattice sites is $N=L^d$. The above energy fluctuation density is proportional to the quantum Fisher information of the operator $\mathcal{O}$ whose scaling has already been studied~\cite{Hauke2016}. Performing a scale transformation $L\to L'=L/\lambda, l\to l'=l/\lambda,m\to m'/\lambda$, scaling operators follow $\mathcal{O}_l \to \lambda^{-\Delta_\mathcal{O}} \mathcal{O}_{l'}$ such that one gets:
\be
	\Delta \epsilon^2 \to \lambda^{d-2\Delta_\mathcal{O}} \frac{1}{L'^d} \sum_{l'm'} \langle \mathcal{O}_{l'} \mathcal{O}_{m'} \rangle.
\ee
Here, it has been assumed that the ground state, where this object is evaluated, is scale invariant. Thus, the energy fluctuations transform as
\be
	\Delta \epsilon^2 \to \lambda^{d-2\Delta_\mathcal{O}} \Delta \epsilon^2,
\ee
Taking into account also the nonuniversal short-range contributions $C$ one obtains at the quantum critical point for a system of finite size:
\be
	\Delta \epsilon^2(L) = \lambda^{d-2\Delta_\mathcal{O}} \varphi (L/\lambda) + C,
\ee
with $\varphi$ encoding the long-range correlations of the operator $\mathcal{O}$. For a system of finite size $L$ the scaling transformation can be continued up to the point $\lambda \sim L$ where the transformation has to be stopped and one obtains:
\be
	\Delta \epsilon^2(L) = L^{d-2\Delta_\mathcal{O}} A + C,
\ee
which is the result presented in the main text. If $d>2\Delta_\mathcal{O}$ the universal contribution is not intensive and dominates over $C$ which is intensive. When $d<2\Delta_\mathcal{O}$ on the other hand, the nonuniversal short-range contributions dominate and the $\Delta \epsilon^2(L)$ stays intensive.

\section{Dynamical quantum phase transitions in the XXZ chain}
\label{app:sec:xxz}

In this part of the supplementary material, numerical data for a further model will be presented supporting that the observations presented in the main text are not tied to the particular model systems bu rather appear more general. The model considered is:
\be
	H(\Delta) = J \sum_{l=1}^{N-1} \Big[ S_l^x S_{l+1}^x + S_l^y S_{l+1}^y + \Delta S_l^z S_{l+1}^z \Big],
\ee
where $S_l^\alpha$, $\alpha=x,y,z$, denotes spin-$1/2$ operators on lattice site $l=1,\dots,N$ with $N$ the system size. Again, open boundary conditions are used. This model exhibits a quantum phase transition at $\Delta=1$ separating an antiferromagnetic phase for $\Delta > 1$ from a Luttinger liquid phase for $\Delta < 1$. The respective order parameter $\mathcal{O} = \mathcal{M}_s$ is the staggered magnetization $\mathcal{M}_s = \sum_l (-1)^l S_l^z$. According to the protocol studied in the main text, the system is initially prepared in the ground state of $H_0=H(\Delta=1)$ at the quantum critical point. Then, the system is quenched by the order parameter such that the dynamics in the system are driven by a Hamiltonian $H=H_0+h\mathcal{O}$ with $h$ the strength of the weak staggered field.

\begin{figure}
\centering
\includegraphics[width=0.5\columnwidth]{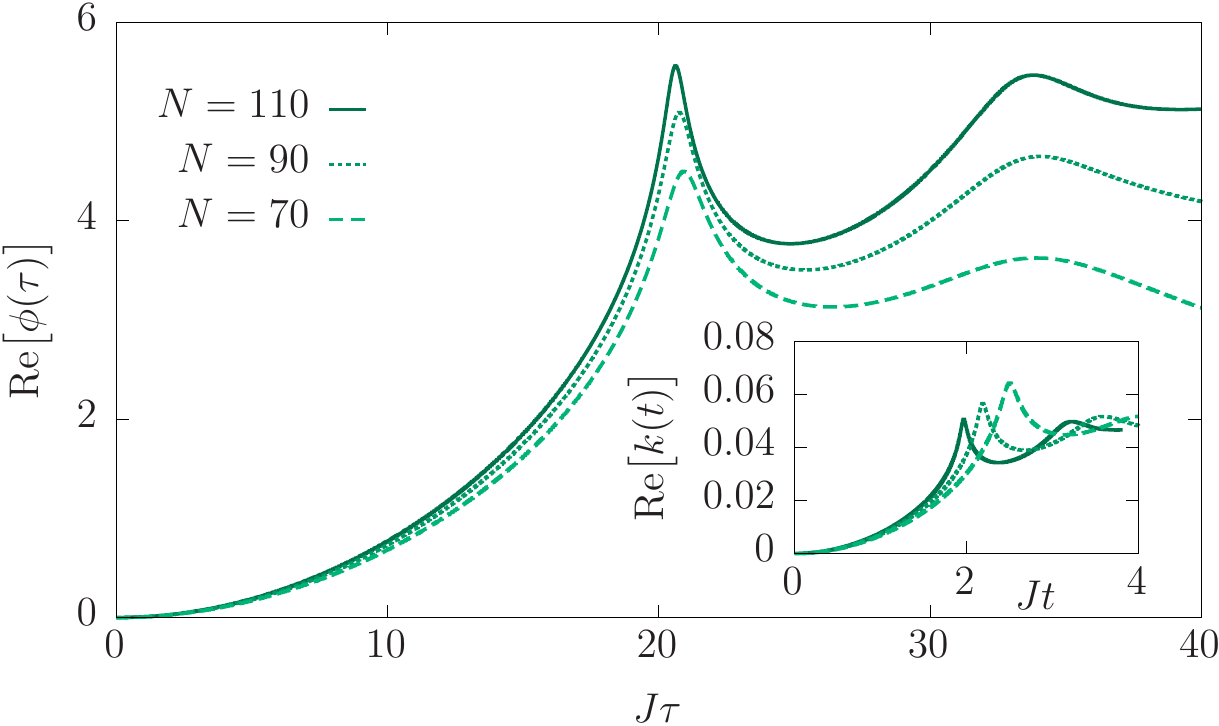}
\caption{Dynamics of the rescaled cumulant generating function $\phi(\tau)$ in the one-dimensional XXZ chain for different system sizes $N$ at a staggered field strength $h/J=0.1$. For comparison, the inset shows the cumulant generating function density $k(t)=K(t)/N$ before rescaling.}
\label{fig:phi_XXZ}
\end{figure}

\begin{figure}
\centering
\includegraphics[width=0.5\columnwidth]{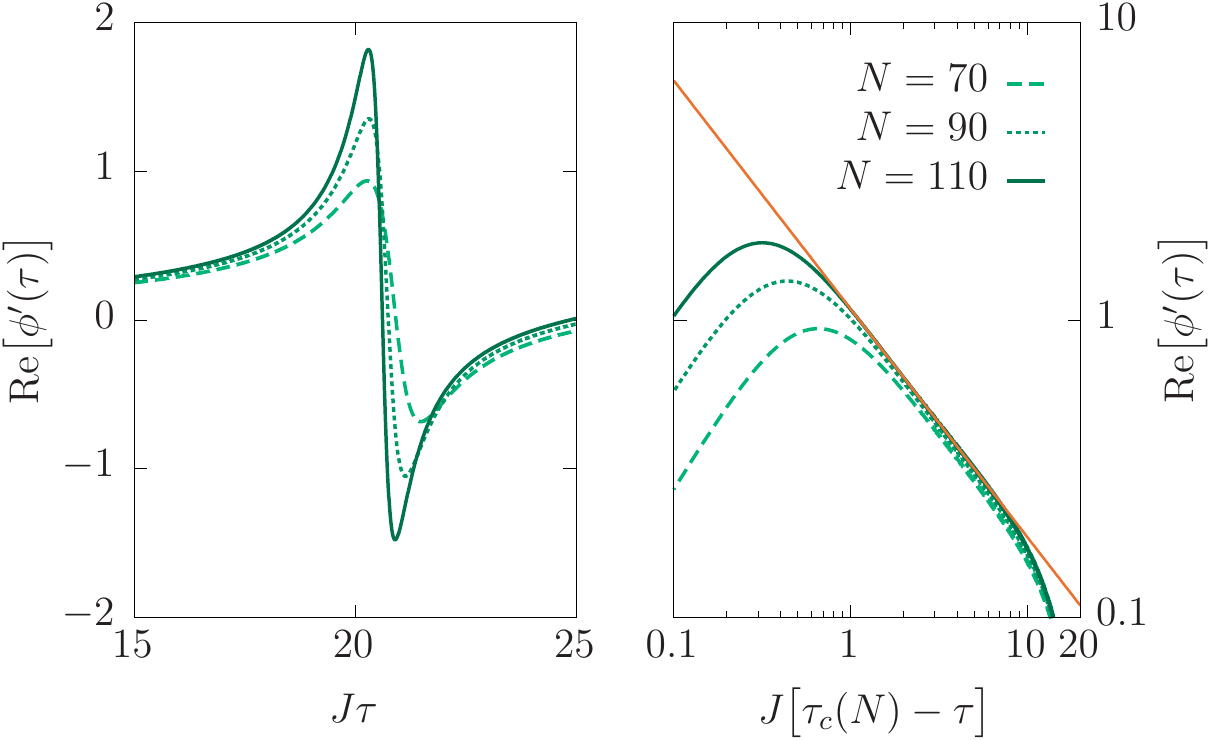}
\caption{Dynamical quantum phase transition in the XXZ chain at a staggered magnetic field strength $h/J = 0.1$.
(left) Real part of derivative $\phi'(\tau) = \partial_\tau \phi(\tau)$ for different system sizes. Around $\tau = \tau_c \approx 20/J$ one observes a sharp peak which becomes more pronounced for larger system sizes. (right) Focus onto the vicinity of $\tau_c$ on a double logarithmic scale indicating a power-law divergence for increasing system size. The algebraic fit $|\tau_c(N) - \tau|^{-\alpha}$ to the curve at N = 110, with
$\tau_c(N)$ the finite-size pseudo-critical point, gives an exponent $\alpha=0.77(5)$.
}
\label{fig:dphi_XXZ}
\end{figure}

In Fig.~\ref{fig:phi_XXZ} the dynamics of the rescaled cumulant generating function $\phi(\tau)$ is shown where the exponents $a=0$ and $b=1/2$ have been used. As for the numerical data in the main text, the iTensor library~\cite{iTensor} has been utilized for the simulations with a bond dimension $\chi=90$ and a time step $\Delta t = 4 \cdot 10^{-4}/\sqrt{N}$. It has been checked that the data has converged both concerning $\chi$ and $\Delta t$. In comparison to the quantum Ising model finite-size effects are much stronger for the studied XXZ chain which might be traced back to the Kosterlitz-Thouless nature of the quantum critical point. Moreover, in Fig.~\ref{fig:dphi_XXZ} numerical data for the derivative $\phi'(\tau)$ of the rescaled cumulant generating function is presented. For increasing system size, the structure around $J\tau\approx 20$ becomes sharper indicating a nonanalytic behavior in the thermodynamic limit. Indeed, on a double-logarithmic scale the data is consistent with an algebraic divergence.

%%%%%%%%%%%%%%%%%%%%%%%%%%%%%%%%%%%%%%%%%%%%%%%%%%%%%%%%%%%%%%%%%%%%%%%%%%

%\bibliography{literature}

%%%%%%%%%%%%%%%%%%%%%%%%%%%%%%%%%%%%%%%%%%%%%%%%%%%%%%%%%%%%%%%%%%%%%%%%%%

\end{document}